\documentclass[a4paper,10pt]{llncs}

\usepackage{ifpdf}				
\ifpdf
	\usepackage[final,pdftex]{graphicx} 	
 	\usepackage[pdftex,pagebackref=false,hyperindex=true]{hyperref} 	
\else
	\usepackage[final,dvips]{graphicx} 	
	\usepackage[dvips,pagebackref=false,hyperindex=true]{hyperref}  	
\fi

\usepackage{amssymb} 

\title{Generate country-scale networks of interaction from scattered statistics}

\author{Samuel Thiriot\inst{1,2} \thanks{This work was partially funded by grant CIFRE 993/2005 from the French National Association for Research and Technology (ANRT).} \and Jean-Daniel Kant\inst{1}}

\institute{Computer Science Laboratory - University of Paris 6 (LIP6), France \and France T\'el\'ecom R\&D - Orange Labs \\ 
\email{Samuel.Thiriot@lip6.fr, Jean-Daniel.Kant@lip6.fr}}

\begin{document}

\bibliographystyle{splncs} 

\maketitle

\begin{abstract}
It is common to define the structure of interactions among a population of agents by a network. Most of agent-based models were shown highly sensitive to that network, so the relevance of simulation results directely depends on the descriptive power of that network. When studying social dynamics in large populations, that network cannot be collected, and is rather generated by algorithms which aim to fit general properties of social networks. However, more precise data is available at a country scale in the form of socio-demographic studies, census or sociological studies. These ``scattered statistics'' provide rich information, especially on agents' attributes, similar properties of tied agents and affiliations. \newline
In this paper, we propose a \textit{generic methodology} to bring up together these scattered statistics with bayesian networks. We explain how to \textit{generate a population of heterogeneous agents}, and how to create links by using both scattered statistics and knowledge on social selection processes. The methodology is illustrated by generating an interaction network for rural Kenya which includes familial structure, colleagues and friendship constrained given field studies and statistics.
\end{abstract}

\section{Context \& problematic}

\subsection{Problematic}
The principle of agent-based models is to reproduce collective dynamics from local interactions. So in any agent-based simulation, the modeler requires a descriptive model of interactions in a population. As these relationships are relatively stable \cite{samuel_thiriot:bib_psycho:wasserman_1994_1}, it became common to represent them using the \textit{social network metaphor}: the structure of interactions is represented by a graph $G(\mathcal{A},\mathcal{L})$, with $\mathcal{A}$ the population of agents and $\mathcal{L}$ the links between these agents. That structure was shown to have a dramatic influence on the dynamics of various agent-based models. The direct consequence is that \textit{the descriptive power of the structure determines the relevance of simulation results}. To ensure that the generated relationships network is descriptive, it should be studied as a modeling problematic: the structure of relationships should be a simplification of social interactions, and comply with knowledge on the modeled population.

While the interaction network can be collected by interview when the population is small, such a data collecting becomes intractable for larger populations. Hence, a lot of models deal with \textit{country-scale populations}, including models of opinion dynamics, virus propagation or information dynamics. We are ourselves interested in modeling diffusion of innovations \cite{samuel_thiriot:bib_perso:thiriot_2008_2}, and propagation of information about these innovations \cite{samuel_thiriot:bib_perso:thiriot_2008_1}. In that field, the lack of descriptive model of interactions was pointed out as one fondamental limitation of models \cite{samuel_thiriot:bib_customer_value:rogers_2003_1}. It is common to use \textit{network generators} to describe such a large population. A network generator is an algorithm which, given several parameters, generates networks compliant with one or more properties observed in real networks. 

Ideally, a network generator should satisfy the following requirements (noted R). \textbf{(R1) generate models of large populations}, in order to improve descriptive power of agent-based models for large-scale simulations. \textbf{(R2) The kinds of relationships linking two agents should be represented}, because interactions don't occur in the same way across different relationships. For instance, finding a work was shown more efficient when activating so-called ``weak ties'' (for instance far family) \cite{samuel_thiriot:bib_psycho:granovetter_1973_1}. \textbf{(R3) Attributes of agents should be detailed} in the network of interactions. That's justified by three main reasons. First, \textit{attributes of individuals influence individual judgment and decision-making} (\textit{e.g.} in diffusion of innovations \cite{samuel_thiriot:bib_customer_value:rogers_2003_1}), so they should be made available in the model. Secondly, \textit{attributes influence the frequency or nature of interaction}: spatial distance reduces frequency of exchanges, differences in ethnicity and interests lower the normative influence, etc. Third, it was shown (as explained in \ref{indoc:evidence}) that \textit{individual characteristics determine the choice of acquaintances} of an agent, so agents' attributes should be took into account during network generation. 

\subsection{Key findings for social networks\label{indoc:evidence}}

Decades of research in social networks highlighted several key findings which are today widely accepted. A stream of research explored \textit{social selection processes} \cite{samuel_thiriot:bib_psycho:robins_2001_1} to understand \textit{how agents create ties}. It appeared that individuals exhibit a strong tendency to create relationships with people sharing similar characteristics (\textit{homophily}) \cite{samuel_thiriot:bib_psycho:mcpherson_2001_1}. Two individuals sharing a common \textit{affiliation} (event, project or workplace) also have more chances to tie and interact frequently \cite{samuel_thiriot:bib_psycho:wasserman_1994_1}. \textit{Transitivity} states that two individuals sharing a common acquaintance are more probably connected together; actually they have more chances to meet together and to create a tie because of a common friend. These observations are no more questionned, so \textbf{(R4) a relevant network of interactions should comply with these processes of social selection}. \newline
The stream of social network analysis also described \textit{statistical properties shared by social networks}, including a surprisingly short average distance between individuals, a high clustering rate (that is, it exists groups or communities in which individuals are strongly interconnected), and a low density \cite{samuel_thiriot:bib_psycho:wasserman_1994_1}. A power-law distribution of degrees was also observed in various datasets (most individuals have few acquaintances, while few have a high degree). It was explained by the so-called preferential attachment principle, which states than new individuals in a network connect more probably with nodes having already a high degree.

Beyond these general properties of social networks, each national institute of statistics publish detailed data for its country. Statistics describe \textit{who the individuals are} by quantifying characteristics on gender, age, ethnicity, socioeconomic class, incomes, marital status, etc. They also study \textit{what people do}: working or not, kind of activity, participation to associative life, sport, etc. These activities can often be interpreted as \textit{affiliations}, with detailed information on agents which are part of the institution (common characteristics of workers like educational level or socioeconomical class, as well as geographical location) and on the affiliations themselves (size, location). More qualitative knowledge also exists on the structure of families, as well as statistics on number of children or household composition. When that kind of data was not collected at a large scale, it is still available from field studies focused on more precise phenomena. As an illustration for this paper, we choose to model social relationships in rural Kenya, for which we have no information from our own. Demographical statistics \cite{samuel_thiriot:bib_customer_value:kdhs_2003_3}, sociological studies on the structure of families (\textit{e.g.} \cite{samuel_thiriot:bib_customer_value:mburugu_2004_1}) and field studies on diffusion of contraceptive use \cite{samuel_thiriot:bib_customer_value:watkins_1995_2,samuel_thiriot:bib_customer_value:rutenberg_1997_1} constitute as many sources of information. Surprisingly, no network generator uses these scattered statistics. However, they constitute an appreciable part of knowledge on the structure of interactions. We claim that these \textbf{(R5) scattered statistics should be taken into account while generating a network of interactions}.

\subsection{Existing models\label{indoc:models}}
%
The most used generators for agent-based models are small-world networks and scale-free graphs \cite{samuel_thiriot:bib_sma_simulation:phan_2007_1}. The first generates networks highly clustered with a short average path length, while the second implements the preferential attachment principle. These models, proposed by physicists, generate highly stylized networks in which social relationships between heterogeneous agents become links between nodes. They neither comply with knowledge on social selection processes nor rely on statistics available for a given population. To summarize, they don't satisfy our requirements R2, R3, R4 nor R5.

In the frame of social network analysis \cite{samuel_thiriot:bib_psycho:wasserman_1994_1}, a lot of models were proposed (see \cite{samuel_thiriot:bib_psycho:robins_2001_1} for a synthetic picture). Existence of a link between two agents $a_1,a_2 \in \mathcal{A}$ is considered to be a random variable $L^{a_1,a_2}$ which takes value $1$ if a link exists. Random graphs with attributes generate links given a vector of agents' attributes $Att$: $p(L^{a_1,a_2}=1 | Att(a_1), Att(a_2))$. If one uses only that constraint, a tie between two agents is independent of any other tie with other agents, in contradiction with transitivity evidence. That assumption was removed by markov random graphs \cite{samuel_thiriot:bib_psycho:frank_1986_1} by allowing two links to be dependent if they share a node in common. In that case links are noted as the conditional probability: $p(L^{a_1,a_2}=1 | L^{a_1,a_3}=1, L^{a_3,a_2}=1)$, with $a_1, a_2, a_3 \in \mathcal{A}$. Recent extensions of these models \cite{samuel_thiriot:bib_psycho:robins_2001_1} take into account both links created given agents' attributes and transitive links; to date, they remain limited to one or two attributes \cite{samuel_thiriot:bib_psycho:robins_2001_1}.\newline 
That formalism is powerful enough to describe homophily and transitivity. Its relevance was proved by fitting data collected from small groups. It was also shown \cite{samuel_thiriot:bib_psycho:robins_2001_1} that affiliations or degree of an agent may be considered to be attributes, so the formalism also enables generation of graphs with power-law degree distribution and affiliations. In short, they fullfit R3 and R4. However, they include special parameters which require Ad Hoc collecting of data, so their application remains limited to small groups (opposite to R1). Moreover, in these small groups, it was never necessary to distinguish different kind of relationships, contrarly to R2.
\subsection{Approach}
We propose to use scattered statistics available for the modeled population (R5) to generate more representative networks of interaction at a country scale (R1). The generated relationships network $G(\mathcal{A}, \mathcal{L}, Att, \mathcal{T})$ will detail agents' attributes $Att$ in the population (R3), which are taken into account during network generation. The network will include several kinds of relationships $\mathcal{T}$ (R2), so the user of that network may infer the interaction network given the kind of relationship (\ref{indoc:usage_simu}).
As the model is intended to bring up together several sources of information on a population to parameter the generator, we describe a methodology (\ref{indoc:methodology}) to formalize intuitively knowledge on agents' attributes and links using bayesian networks. The formalism, inspired by markov random graphs, enables representation of the key processes of social selection (R4). Then (\ref{indoc:generation}), we explain how a population of heterogeneous agents can be generated and how the relationships network is created. Insights on the minimum size of population, on detection of statistics discrepancies, and on statistical properties of generated networks are provided in section \ref{indoc:application}.

\section{Methodology\label{indoc:methodology}}

\subsection{Choice of agents' attributes and link types}
\paragraph*{Step 1}
The modeler should first \textbf{define the types of social links} $\mathcal{T}$ he wants to represent in the relationships network. $\mathcal{T}$ enumerates links leading potentially to different interactions in the model, or which are created by different processes. As proposed previously in markov random graphs \cite{samuel_thiriot:bib_psycho:frank_1986_1}, some kinds of relationships can be generated given agents' attributes $\mathcal{T}^{att}$, while others are created by transitivity $\mathcal{T}^{trans}$. In the example of Kenya, we choose to represent links leading to interaction about contraceptive use. Field studies indicate that spouses discuss that topic, that advices of parents have a normative influence, and that women retrieve information from friends, siblings and colleagues \cite{samuel_thiriot:bib_customer_value:rutenberg_1997_1}. We define links created given agents' attributes $\mathcal{T}^{att}=\{$\textit{spouses, motherOf, colleagues, friends}$\}$. Other links are created by transitivity, because they involve more than two agents and can be created given already created links: $\mathcal{T}^{trans}=\{$\textit{fatherOf, siblings}$\}$. 

\paragraph*{Step 2}
Next the modeler has to \textbf{select agents' attributes} $Att$  which are known - or supposed - to influence probability of a link to be created. Of course, that selection is done given available data and the purpose of the agent-based model. Typically agents' attributes will contain \textit{socio-demographic characteristics} (age, gender, socioeconomic class, ethnicity, etc.) and \textit{places were the agents have frequent interactions} given these characteristics (going to school, frequenting a workplace, etc.). We also assume the number of links to create for an agent for each kind of relationship to be an attribute:  $\forall t \in \mathcal{T}$, $RC^{a}_t \in Att(a)$. The choice of including the number of links as an attribute could seem counter-intuitive, because it was often considered to be an independent density parameter \cite{samuel_thiriot:bib_psycho:robins_2001_1}. That choice is justified by the following reasons: (i) The number of links per agent is available from statistics, and varies across kinds of relationships (ii) \textit{the number of links is strongly correlated to other agents' attributes}; for instance the number of children of a wife depends on its age. (iii) \textit{the number of links is often considered to be an explanatory variable} for the individual decision-making process, so it should be made available as an attribute. As example, contraceptive adoption increases with the number of children of a mother \cite{samuel_thiriot:bib_customer_value:watkins_1995_2}. \newline
In the example of Kenya, attributes \textit{married}, \textit{age} and \textit{gender} are required for nearly all kinds of links. As field studies indicate that most discussions take place during quotidian activities \cite{samuel_thiriot:bib_customer_value:watkins_1995_2}, we added the variable \textit{work}. \textit{Spatial location} is required because spouses always live in the same place, as do young children with their mother.

\subsection{Formalization based on bayesian networks}

\paragraph*{Step 3: represent agents' attributes using bayesian networks.\label{indoc:step2}}

\begin{figure}[t]
\centering
\includegraphics{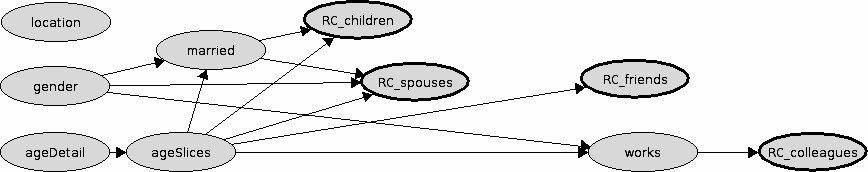}
\caption{Attributes bayesian network used to describe interdependencies between Kenyan socio-demographic attributes. Nodes in bold are the number of links to create for each link type.}
\label{fig:bayesian_network_1}
\end{figure} 

Attributes of individuals in a real population are strongly interdependent: marital status depends on age and gender, socioeconomic class is highly correlated with location and education level, etc. Generating a population of agents in which attributes of agents comply with statistical interdependences of individual characteristics requires a relevant modeling of these dependencies, generic enough to be used with any kind of data. Hence, data available for a population is often presented as statistics linking one attribute with another. For instance, the number of children per woman is provided given marital status and age \cite[p.~57]{samuel_thiriot:bib_customer_value:kdhs_2003_3}. That kind of statistic can be translated, without loss of generality, to conditional probabilities, like $p\left(RC^{a}_{motherOf}=\{0...10\} | age(a), maritalStatus(a)\right)$. In that viewpoint, \textit{attributes of agents are considered to be random variables}. We propose to use a bayesian network  \cite{samuel_thiriot:bib_model_decision:jensen_1996_1} (BN)
, named \textit{agent BN} in this methodology, to formalize these interdependencies. Each agent attribute in $Att$ is represented by a variable in the BN. The domain of a variable defines the values the attribute can take. For instance in graph \ref{fig:bayesian_network_1}, variable \textit{gender} has domain $D^{gender}=\{ male, female \}$, $D^{married}=\{yes,no\}$, and $D^{RC\_motherOf}=\{0...10\}$. Root variables define initial probabilities. In Fig~\ref{fig:bayesian_network_1}, initial probabilities for variable \textit{ageDetail} define the probability for an individual picked up randomly in the population to have a given age; that probability is available from the age pyramid of the target population. A directed link between two variables $V_1 \rightarrow V_2$ means that $V_2$ probabilities can be calculed using its parents, and only its parents. $V_2$ embodies a conditional probability table representing the probability to take each value $D^{V_2}$ given all the possible values in the domains of its parents (here, $V_1$). No link means that variables are assumed independent. That don't means that variables are independent in reality, but rather represent our lack of knowledge (or our willingness to simplify that knowledge) of that dependence. 

In our application to Kenya, probabilities in the agent BN depicted in Fig.~\ref{fig:bayesian_network_1} come from the US Census Bureau, from the Kenya demographic and health survey \cite{samuel_thiriot:bib_customer_value:kdhs_2003_3}, and from field studies (e.g. \cite{samuel_thiriot:bib_customer_value:watkins_1995_2,samuel_thiriot:bib_customer_value:rutenberg_1997_1}). Note that we used \textit{convenience variables} to simplify formalization of data: in agent~BN~\ref{fig:bayesian_network_1}, variable \textit{ageSlices} simplifies the detailed age to 5-year slices, which are often used in published statistics. Another benefit of BN is to \textit{highlight evident discrepancies in data}. For instance, a social scientist will immediately note in (Fig.~\ref{fig:bayesian_network_1}) the absence of link between gender and age, while the age pyramid in most of countries shows significant differences between genders (indeed, Kenya is a particular case of symmetrical age pyramid). 
%
%
%
\begin{figure}[t]
\centering
\includegraphics{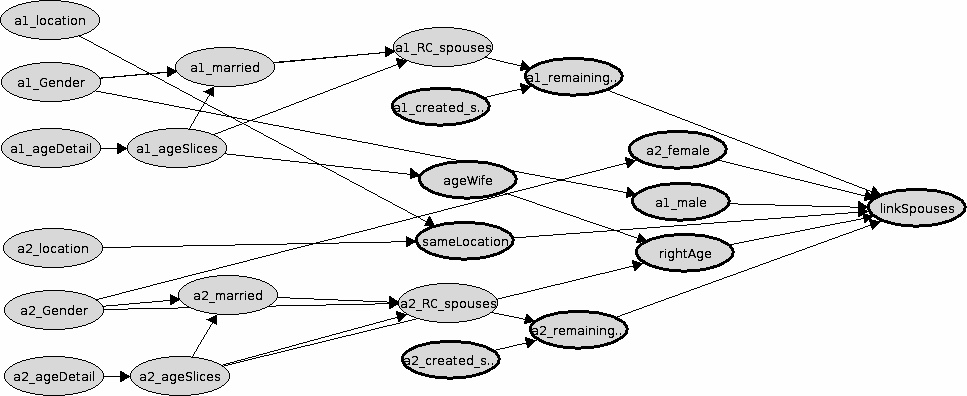}
\caption{Matching bayesian network for link type \textit{spouses}. On the left, the agent BN for agents 1 and 2.}
\label{fig:bn_appariement_1}
\end{figure} 
\paragraph*{Step 4: represent links probability using bayesian networks.}
Links created given attributes $\mathcal{T}^{Att}$ are defined by $p( L^{a_1,a_2}_t=1 | Att(a_1), Att(a_2) )$. They can be used to represent a large range of phenomena, including \textit{homophily}, \textit{affiliation}, \textit{preferential attachment}, or spatialization. As that probability is conditional, it can also be represented by a \textit{matching BN}, for which an example for relationship \textit{spouses} in Kenya is depicted in Fig.~\ref{fig:bn_appariement_1}. In that matching BN, one can recognize two instances of the agent BN (on the left) representing attributes of two different agents of the population. On the right, a special node with domain $\{yes,no\}$ defines if a link can be created between these agents. Nodes in bold define constraints on linking. In the BN in Fig.~\ref{fig:bn_appariement_1}, we define arbitrarily that agent 1 is male and agent 2 female (wedding in Kenya is heterosexual). Node \textit{ageWife} projects the probable age of the first wife of man described on top (on average 10 years younger), and variable \textit{rightAge} ensures by an identity probability table that agent 2 complies with that age. The node \textit{sameLocation} takes value \textit{yes} only if both agents live in the same location. The final variable ``linkSpouses'', which determines if two agents can be linked together, takes values ``yes'' only if all of its parents are themselves to ``yes''. Note the nodes \textit{a1\_created\_spouses} and \textit{a1\_remaining\_spouses}, which ensure that we will only create as many links of type $t$ as required by $RC_t^{a_1}$ and $RC_t^{a_2}$, but no more, so a wife will exactly be tied with one husband. Other links in $\mathcal{T}^{Att}$ are defined in the same way: friends have probably the same age and live probably in the same town, mothers are linked to children whom age is compliant with their age (and live always in the same location if children are young), and colleagues are defined as agent sharing the same activity in the same location.

Links created by transitivity $\mathcal{T}^{trans}$ are also random variables, and are noted: $p(L^{a_1,a_2}_{t_1}=1 | L^{a_1,a_3}_{t_2}=1, L^{a_3,a_2}_{t_3}=1)$, with $a_1, a_2, a_3 \in \mathcal{A}$, $t_1 \in \mathcal{T}^{trans}$, $t_2,t_3 \in \mathcal{T}$. That formalism is quite intuitive and will not be more detailed here. In our example, we define by transitivity the link ``fatherOf'' with  $p(L^{a_1,a_2}_{fatherOf}=1 | L^{a_1,a_3}_{motherOf}=1, L^{a_3,a_2}_{spouses}=1)=1$ (only transitivity enable description of father-children links; it could not be described as a matching BN, because children of a man have to be the same than children of its wives). In the same way, siblings are created by transitivity across mother and father links. With a lower probability, friendships links are created by transitivity between friends.

\section{Generation of the graph\label{indoc:generation}}

\subsection{Generation of an heterogeneous population}

\begin{figure}[t]
\centering
\begin{tabular*}{12cm}{c}
\vspace{0.5em}\includegraphics{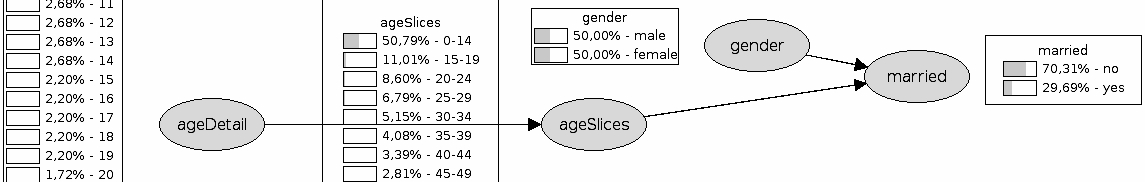} \\ 
\includegraphics{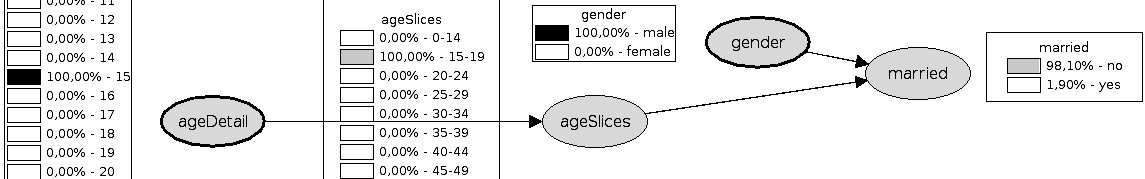}
\end{tabular*}
\caption{\footnotesize Example of evidence propagation when the bayesian network is used to generate agents' attributes. Here monitors (boxes in the figure) display the probabilities for each variable to take every value (note that some of these monitors are truncated). \textit{(top)} probabilities with no evidence \textit{(bottom)} probabilities when evidence is set.}
\label{fig:evidence_propagation}
\end{figure}

All the variables $Att$ in the agent BN will become agents' attributes with the same domain. For each agent $a \in \mathcal{A}$ to create, we generate a \textit{prototype agent}. The process to generate a prototype simply consists in using the agent BN in a generative way: for each variable $V \in Att$ of the agent BN (in the ordinal order, so root variables are processed first), a value $V=v$ is selected randomly in the domain of $V$, given probabilities $p(V=v | parents(V))$ defined in the BN. When value $V=v$ has been chosen, a corresponding piece of evidence $p(V=v)=1$ is put in the BN. Evidence, in the theory of BN, represents a known information. Putting evidence in the BN permits to compute probabilities of child variables given the values of already selected attributes, so the integrity of agent attributes is ensured. For instance in Fig.~\ref{fig:evidence_propagation}, before any piece of evidence \textit{(top)}, the probability for someone randomly picked up in the population to be married is 29.69\%. When attributes $ageDetail$  and $gender$ have been randomly put to $15$ and $male$, and used as evidence, posterior probability for the current agent to be married falls to 1.90\%. When all the agents are generated that way, the statistical distribution of their attributes complies with the distribution described by the BN.

\subsection{Creation of links}

Now that all agents were created in the population $\mathcal{A}$, each agent having its attributes $Att(a)$ defined, we have to link them using the matching BN. For each kind of relationship $t \in \mathcal{T}^{Att}$, we constraint the matching BN for $t$ by providing \textit{evidence for link creation}: as our aim is to link together agents with link $t$, we set evidence on variable $p(link\_spouses=yes)=1$. Given that evidence, probabilities for attributes of $a_1$ and $a_2$ are updated, and some probabilities in attributes' domains fall to zero. For instance, in the case of ``spouses'' link, probability of agents 1 and 2 to be younger than 15 years falls to zero; they also cannot have ``married=no''. In other words, probabilities in the matching BN given evidence of link creation designate two sets of \textit{candidates for linking} $\mathcal{C}^t_1$ and $\mathcal{C}^t_2$. The matching process will remain limited to these sets. 
Then, we iterate across candidates $\mathcal{C}^t_1$ and select randomly has many acquaintances among $\mathcal{C}^t_2$ as required by $RC^{a_1}_t$. For each agent $a_1 \in \mathcal{C}^t_1$, we load its attributes and use them as pieces of evidence in the BN. After a run of the inference engine, the probabilities for agent 2 define a restricted set $\mathcal{C}^t_2 | Att(a_1) \subset \mathcal{C}^t_2$ of candidates for linking \textit{given agent 1 attributes}. In our application to Kenya, for link type $spouses$, $C_1$ is the set of husbands and $C_2$ the set of wives. When one chooses an agent $a_1$, given the constraints on matching, the set $\mathcal{C}_2 | Att(a_1)$ limits $\mathcal{C}_2$ to wives which live in the same location than $a_1$. Selecting a candidate is made by generating a prototype agent as explained before; if that prototype cannot be found, a fallback solution consists in picking up randomly one agent in $\mathcal{C}^t_2 | Att(a_1)$ (note that the fallback solution can bias statistical distribution in the population; in our example it is possible to link a husband with an older wife). When no fallback solution can be found, that is when $\mathcal{C}_2 | Att(a_1) = \{ \varnothing\}$, agent $a_1$ remains orphan, but will never be tied with a incompatible agent (in our example, no man will be said married with a non married or too young wife). These errors will be studied in \ref{indoc:errors}.

After having processed all link types defined by matching BN, transitive links are created using the probabilities $p(L^{a_1,a_2}_{t_1} | L^{a_1,a_3}_{t_2}, L^{a_3,a_2}_{t_3})$ formalized in step~2.

\section{Generated network\label{indoc:application}}

\subsection{Usage for social simulation\label{indoc:usage_simu}}

\begin{figure}[t]
\includegraphics[width=12cm]{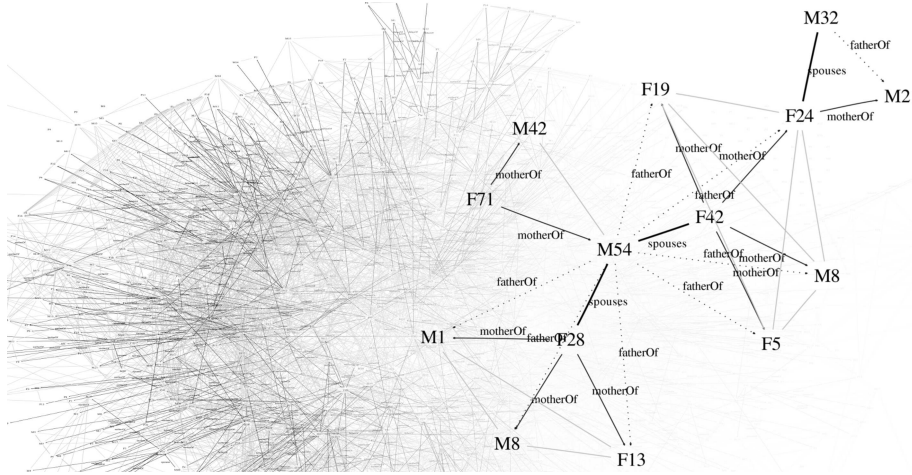} 
\caption{\textit{(left)} generated relationships network \textit{(right)} zoom in one agent}
\label{fig:result}
\end{figure} 

The resulting graph $G\left( \mathcal{A}, \mathcal{L}, Att, \mathcal{T} \right)$ includes links $\mathcal{L}$ of different kinds ($\mathcal{L} = \bigcup_{t \in \mathcal{T}} \mathcal{L}^t$), and provides the values of agent attributes $Att$ for any agent in the population $\mathcal{A}$.
The structure of relationships described by the generated network depends obviously on agent BN and matching BN provided by the modeler as parameters. In our application to Kenya, the population covers the whole age pyramid, and describes attributes depicted in Fig.~\ref{fig:bayesian_network_1}. Moreover, as shown in Fig.~\ref{fig:result}, each agent is positionned in its familial environment; agent M54 (for Male, 54 years) is married with two wives F28 and F42, and has 7 children, including one daughter F24 which is herself married and mother. He is also tied with its own mother F71 and brother M42, but not with its father - probably because this one is not in the age pyramid (no more alive). He his also tied with \textit{colleagues} and \textit{friends} (not represented in that figure to improve lisibility). That structure is described at the scale of the 50,000 agents depicted in Fig~\ref{fig:result} \textit{(left)}. \newline
To use that network for simulation, the modeler may simply define probability to interact given the kind of relationship: $\forall t \in \mathcal{T}$, $p^t_{interact}$, so  $p_{interact}(a_1,a_2 | L^{a_1,a_2}_t) = p^t_{interact}$. He may also choose a finer granularity by defining the probability of interaction given attributes $Att$, for instance to represent the fact that spatial distance decrease probability of interaction. In that illustration, we focus on interactions about contraceptive use \cite{samuel_thiriot:bib_customer_value:rutenberg_1997_1}. In our case, no interaction occurs across links between young children and their parents. As the topic of contraceptive use is sensitive in Kenya, probabilities of discussion between spouses are low, as between a mother and its own parents. In fact, women which are still fertile and are concerned by the topic discuss mainly with their female friends, and often with their brothers-in-law (link sibling). The resulting network of interactions is a network in which ties are weighted by probabilities; it is by far sparse than the network of relationships.

\subsection{Errors and statistical properties\label{indoc:errors}}
\begin{figure}[t]
\begin{tabular*}{\linewidth}{c c}
\includegraphics[width=6cm]{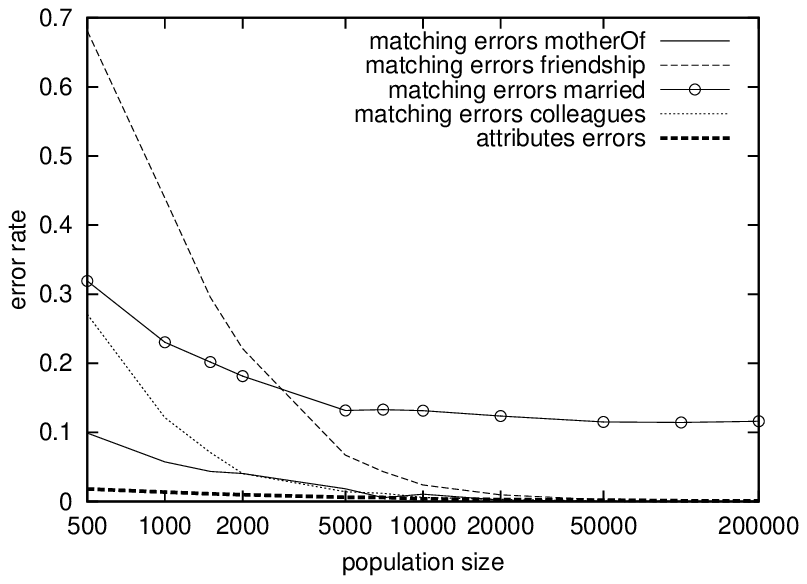} & 
\includegraphics[width=6cm]{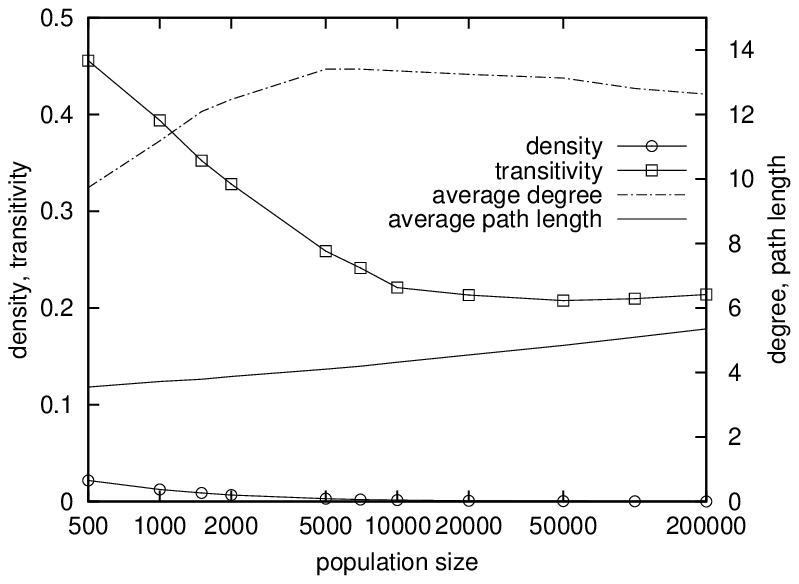} \\
\end{tabular*}
\caption{\textit{(left)} Error rate given population size. \textit{(right)} Statistical properties of the generated relationship network}
\label{fig:errors}
\end{figure} 

While BN describe a theoretical population using continuous probabilities, we generate a discrete population and link agents only when a suitable candidate exists. That limitation necessarily leads to a bias in the statistical properties of the population. Two kinds of errors may appear during generation. \textit{Errors on statistical distribution} appear because the generated population $\mathcal{A}$ is not large enough given the combinatories of attributes' values described by the agent BN. These errors are measured by learning the agent BN on data, and quantified as the average difference between theoretical and measured probability. As shown in Fig.~\ref{fig:errors} \textit{(left)}, these errors (bottom curve) remain low and are negatively correlated to the population size. \textit{Errors on matching} appear when no candidate was found to link several agents, and are quantified as the rate of the total number of links required by $RC_t$ on the number of created ties. When that error rate remains low, and decreases when the population size increases, errors are only due to the discrete nature of agents: it will always exist some agents which could not be connected because their theoretical peer was not created. As shown in Fig.~\ref{fig:errors}, that error rate drops quickly above a given population size. Given our parameters, a population of 5,000 agents is a minimum to reduce errors. Above 10,000 agents, no more significant improvement appears. When the matching error rate remains high when the population size increases, it means that agent BN and/or matching BN are incompatible. In Fig.~\ref{fig:errors}, curve for link \textit{married} shows that the number of wives per men is not compatible with the proportion of married wives. In that case statistics (or assumptions) used to build BN should be checked and corrected. 

Figure~\ref{fig:errors} \textit{(right)} depicts the evolution of statistical properties of the relationships network. Density is low (under 0.01). Transitivity (sometimes called clustering rate) is high, and becomes stable above the 10,000 agents threshold. Links defined across spatial locations (for family, work, and with low probability friendship) play the role of shortcuts, so the average path length in the model grows very slowly (around 4.8), exhibiting the so-called ``small-world'' property. The average degree is in theory defined by attributes $RC_t$. In fact, it is only reached when all the required links are created (above 10,000 agents), then sticks to its theoretical value. \newline
At evidence, it exists a minimum population size to satisfy constraints defined by matching BN. The more matching BN are constraining, the higher the threshold. Above that threshold, statistical properties remain remarkably stable.

\section{Discussion\label{indoc:discussion}}

In this paper, we proposed a methodology to formalize various statistics available for the population (R5) to generate a simplified network of relationships at a country scale (R1). The resulting network of relationships includes agents' attributes (R3) and different kinds of relationships (R2), so the modeler can define with more precision if interaction takes place. The network exhibits a high clustering rate, low density and a low average path length. Formalism enables modelers to comply with evidence on social selection processes (R4) like affiliation, homophily and transitivity. We illustrated the methodology by generating a network of relationships for rural Kenya in which socio-demographic studies, sociological findings, and qualitative observations on affiliation are put together to reproduce familial, work and friendship relationships.\newline
The choice of bayesian networks to formalize scattered statistics make the fusion of different statistical sources more intuitive, so any social scientist can use the generator. BN also facilitate generation of the heterogeneous population of agents and the creation of links between these agents. We plan to publish soon the software which implements the generator. 

The purpose of that methodology is to generate a \textit{model} of relationships in a population. So, the generated graph is only a \textit{simplification of real relationships given available data}, and don't target the same precision than models at a smaller scale. However, the network generated is rooted in reality by using available statistics and observations of that population. In some way, we hope it fills the gap between models from social scientists (highly descriptive, but limited to small populations) and generators from physicists (generate large populations with a low descriptive power). \newline
We decided to illustrate this paper with social interactions in rural Kenya because of the relative simplicity of its social structure. The next step is to model more complex populations like France  (more affiliations, socioeconomic classes, attributes). Our agenda of research also includes the \textit{investigation of dynamics} supported by generated networks, especially in the frame of information diffusion, and the \textit{formal analysis of the properties of generated networks}.


\bibliography{../../commons/biblio_bibtex/library}

\end{document}